\documentclass[showkeys,notitlepage,superscriptaddress,floatfix,prd,10pt,aps]{revtex4-1}

\usepackage[utf8]{inputenc}
\usepackage[colorlinks=true,citecolor=blue,linkcolor=blue]{hyperref}
\usepackage[normalem]{ulem}
\usepackage{url}
\usepackage{graphicx,wrapfig,float,slashed,cancel}
\usepackage{amsmath,amssymb,epsfig,graphicx,xcolor,stmaryrd}
\usepackage{bbold} 
\usepackage{bm}
\usepackage{enumitem}
\usepackage{multirow}

\usepackage{physics}
\usepackage{orcidlink}

\definecolor{darkblue}{RGB}{1, 90, 173}

\begin{document}
	
	
\title{QCD sum rule study of the tensor $\Delta^0\Delta^0$ dibaryon state}
\author{M. Ahmadi\orcidlink{0009-0006-4046-2121}}
\email{masoumehahmadi@ut.ac.ir}
\affiliation{Department of Physics, University of Tehran, North Karegar Avenue, Tehran 14395-547, Iran}
\author{H. Mohseni}
\email{hosseinmohseni@ut.ac.ir}
\affiliation{Department of Physics, University of Tehran, North Karegar Avenue, Tehran 14395-547, Iran}
\author{K. Azizi\orcidlink{0000-0003-3741-2167}}
\email{kazem.azizi@ut.ac.ir}
\thanks{Corresponding author}
\affiliation{Department of Physics, University of Tehran, North Karegar Avenue, Tehran
14395-547, Iran}
\affiliation{Department of Physics, Dogus University, Dudullu-\"{U}mraniye, 34775
Istanbul,  T\"{u}rkiye}

\date{\today}
	
\preprint{}
	
\begin{abstract}

We study the exotic $\Delta^{0}\Delta^{0}$ dibaryon state with quantum numbers $J^{P} = 2^{+}$ within the framework of QCD sum rules. A tensor interpolating current is constructed to determine the mass and residue of this state, including explicit calculations of contributions from quark, gluon, and mixed vacuum condensates up to dimension nine. The numerical analysis yields a mass of 
$m_{\Delta^{0}\Delta^{0}} = 2426^{+101}_{-108}~\mathrm{MeV}$ 
and a residue of 
$f_{\Delta^{0}\Delta^{0}} = \left( 2.30^{+0.49}_{-0.45} \right) \times 10^{-4}~\mathrm{GeV}^{8}$. 
The predicted mass lies about $38~\mathrm{MeV}$ below the $2m_{\Delta^{0}}$ threshold, which indicates the possible existence of a bound configuration. Consequently, the tensor $\Delta^{0}\Delta^{0}$ dibaryon may serve as a promising candidate for future experimental searches.

\end{abstract}

	
\maketitle
	
\renewcommand{\thefootnote}{\#\arabic{footnote}}
\setcounter{footnote}{0}
\section{Introduction}\label{intro} 
The investigation of multiquark states within quantum chromodynamics (QCD) began shortly after the theory was established in the 1970s. A pioneering bag-model calculation from that period ignited interest in these exotic systems and motivated both experimental searches and alternative theoretical approaches. While conventional hadrons, such as mesons (quark--antiquark pairs) and baryons (three-quark states), were well understood within the quark model, the possibility of exotic hadrons, such as tetraquarks (four quarks) and pentaquarks (five quarks), attracted considerable theoretical attention~\cite{Gell-Mann:1964ewy}. Early theoretical models predicted the existence of multiquark states, but experimental confirmation remained elusive for decades.

Significant progress occurred in the 2000s due to advancements in experimental techniques and particle accelerators. In 2003, the Belle Collaboration reported the discovery of the $X(3872)$, widely regarded as the first strong tetraquark candidate. This was followed by further discoveries, including the observation of pentaquark states by the LHCb experiment at CERN, where so-called ``exotic'' hidden-charm $XY\!Z$ tetraquark and $P_c$ pentaquark states were reported. These milestones marked a turning point in multiquark research, opening new avenues for exploring QCD and deepening our understanding of hadron structure and the strong interactions that govern their behavior~\cite{Belle:2003nnu,Swanson:2006st,LHCb:2015yax,Chen:2016qju,Lebed:2016hpi,Olsen:2017bmm}.

Dibaryons, as another important class of multiquark hadrons, pose some of the most complex theoretical and experimental challenges in strong interaction physics. The deuteron, discovered in 1932 by Urey, Brickwedde, and Murphy~\cite{Urey:1932gik}, was the first identified dibaryon. Another intriguing example is the H-dibaryon predicted by Jaffe~\cite{Jaffe:1976yi}, which continues to attract both experimental and theoretical interest. More recently, the $d^*(2380)$ resonance was observed in the total cross section at $\sqrt{s} = 2.37~\mathrm{GeV}$ with $\Gamma \approx 70~\mathrm{MeV}$ and quantum numbers $I(J^P) = 0(3^+)$, in double-pionic fusion reactions reported by the WASA-at-COSY Collaboration~\cite{WASA-at-COSY:2011bjg,WASA-at-COSY:2012seb,Gal:2013dca,WASA-at-COSY:2014dmv,WASA-at-COSY:2014qkg}. Following its observation, numerous studies have examined the $d^*(2380)$, including investigations of its total width~\cite{Dong:2015cxa}, as well as analyses of its mass, binding energy, root-mean-square radius, and electromagnetic form factors in comparison with experimental data~\cite{Dong:2018ryf,Dong:2019zvg}. The $d^*(2380)$ has also been studied using QCD sum rules~\cite{Chen:2014vha}, where three $\Delta$--$\Delta$-like interpolating currents yielded a predicted mass $M_{d^*} = 2.4 \pm 0.2~\mathrm{GeV}$, consistent with the WASA measurement. Further research has investigated its structure and decay properties using a chiral constituent quark model~\cite{Huang:2019lzt}, evaluated the properties of its decuplet states~\cite{Bashkanov:2020lxv}, and explored its possible configurations as a hexaquark state~\cite{Kim:2020rwn}.

Recent investigations have examined the influence of the $d^*(2380)$ hexaquark on the equation of state (EoS) of dense neutron star matter, exploring the implications of its existence for neutron star properties. These studies have shown that including the $d^*$ state notably softens the EoS at high densities, thereby reducing the predicted maximum mass of neutron stars~\cite{Celi:2023gtj,Celi:2025wnc}. Motivated by the possibility of a stable hexaquark $(uuddss)$ as a dark matter candidate, Ref.~\cite{Shahrbaf:2022upc} investigated its compatibility with neutron star observations, focusing on constraints from maximum mass and tidal deformability.

 In parallel, interest in dibaryons has been fueled by proposals to explore them as potential dark matter candidates~\cite{Farrar:2003qy,Farrar:2017eqq,Farrar:2018hac}. Studies on the mass and stability of the H-dibaryon, inspired by Jaffe's original prediction~\cite{Jaffe:1976yi}, have been carried out using various particle physics models. The corrected MIT bag model yields $m_{H} = 2240~\mathrm{MeV}$, placing it just above the $2m_{\Lambda}$ threshold~\cite{Halprin:1982pb}, while a chiral model predicts a substantially lower value of $1130~\mathrm{MeV}$~\cite{Yost:1985mj}.

 Various quark models have also been employed to study the $\Lambda\Lambda$ interaction and its binding energy~\cite{Oka:1986fr,Straub:1988mz,Koike:1989ak}. QCD sum rule calculations give results consistent with Jaffe's initial estimate, indicating a mass in the range $2.0$--$2.4~\mathrm{GeV}$~\cite{Kodama:1994np}. Moreover, some experimental studies have reported positive signals to find a stable dibaryon state~\cite{Khrykin:2000gh,Bashkanov:2008ih}. In particular, Ref.~\cite{Bashkanov:2008ih} reports data indicating the formation of a $\Delta\Delta$ system in the intermediate state, suggesting that the ABC effect is associated with the $(\pi\pi)_{I=L=0}$ channel. Using a simple string model, Ref.~\cite{Vijande:2011im} found that the ground state of $(q^{6})$ remains stable against dissociation into two baryons and predicted a bound state very near the threshold for the $(q^{3}\bar{q}^{3})$ case.

 Further evidence for a bound H-dibaryon comes from lattice QCD simulations. A bound state with $I = 0$, $J = 0$, and quark content $uuddss$ was found at $m_{\pi} \sim 389~\mathrm{MeV}$, with a binding energy of $16.6 \pm 2.1 \pm 4.6~\mathrm{MeV}$, where the first uncertainty is statistical and the second systematic. With more measurements and improved analysis, the binding energy was refined to $B_{H} = 13.2 \pm 1.8 \pm 4.0~\mathrm{MeV}$~\cite{NPLQCD:2010ocs,NPLQCD:2011naw}. Using the baryon--baryon potential method described in Ref.~\cite{Ishii:2006ec}, the HALQCD Collaboration performed simulations on three lattice volumes and at three quark masses, reporting a bound H-dibaryon with a binding energy of $30$--$40~\mathrm{MeV}$ for pion masses between $673$ and $1015~\mathrm{MeV}$~\cite{Inoue:2011nq}. Issues related to Lattice QCD were also investigated using chiral effective field theory at leading order for baryon--baryon interactions, focusing on the $\Xi\Xi$, $\Xi\Sigma$, and $\Xi\Lambda$ channels. This analysis revealed significant effects from SU(3) symmetry breaking, arising mainly from differences among the $\Lambda\Lambda$, $\Sigma\Sigma$, and $\Xi N$ thresholds. At physical masses, this breaking was found to reduce the H-dibaryon binding energy by as much as $60~\mathrm{MeV}$ compared to the SU(3)-symmetric case~\cite{Haidenbauer:2011za}.

 The interaction between the nucleon ($N$) and the Omega baryon ($\Omega$) was investigated in Ref.~\cite{HALQCD:2014okw}, where a bound state was identified with a binding energy of $18.9(5.0)^{+12.1}_{-1.8}~\mathrm{MeV}$. Lattice QCD studies of the $\Omega\Omega$ dibaryon in the ${}^1S_{0}$ channel reported a binding energy of $B_{\Omega\Omega} = 1.6(6)^{+0.7}_{-0.6}~\mathrm{MeV}$, indicating an overall attraction near the unitary regime~\cite{Gongyo:2017fjb}. Following this, analyses of two-particle momentum correlations for $\Omega\Omega$ and $p\Omega$ pairs in relativistic heavy-ion collisions predicted shallow bound states characterized by comparatively large positive scattering lengths in the $J = 0$ $\Omega\Omega$ and $J = 2$ $p\Omega$ channels~\cite{Morita:2019rph}.

 Using QCD sum rules, the exotic $\Omega\Omega$ dibaryons with $J^{P} = 0^{+}$ and $2^{+}$ have been studied in a molecular picture. The predicted scalar state has a mass of $(3.33 \pm 0.22)~\mathrm{GeV}$, suggesting a loosely bound system, while the tensor state at $(3.24 \pm 0.23)~\mathrm{GeV}$ may be more deeply bound~\cite{Chen:2019vdh}. The production of $p\Omega$ (${}^{5}S_{2}$) and $\Omega\Omega$ (${}^{1}S_{0}$) in central Pb+Pb collisions at $\sqrt{s_{NN}} = 2.76~\mathrm{TeV}$ was studied in Ref.~\cite{Pu:2024kfh}, considering two possible structural scenarios: molecular states and compact six-quark states.

 In Ref.~\cite{Junnarkar:2024kwd}, the ground state spectrum of two-flavor heavy quark baryons within the spin-singlet channel was investigated. Moreover, the relationship between the binding energy and the constituent quark masses was explored by varying quark masses across a range from strange to bottom quark masses. The S-wave $\Lambda\Lambda$ and $N\Xi$ systems were examined using lattice QCD~\cite{HALQCD:2019wsz}, with results indicating that the $N\Xi$ interaction lies close to unitarity, whereas the $\Lambda\Lambda$ interaction at low energies does not support a bound state.

 Theoretical studies on the $N$--$\Omega$ interaction have been presented in Refs.~\cite{Morita:2016auo,Haidenbauer:2017sws,Garcilazo:2018gkb,HALQCD:2018qyu}. Additionally, Ref.~\cite{Chen:2021hxs} employed QCD sum rules and reported that an $N\Omega$ dibaryon bound state may exist in the (${}^{5}S_{2}$) channel, with a binding energy of approximately $21~\mathrm{MeV}$. The spectra of potential hidden-bottom and hidden-charm hexaquark states were investigated using the QCD sum rule framework~\cite{Wan:2019ake}. This analysis suggests the possible existence of two baryonium states in the bottom-quark sector, with estimated masses of $11.84 \pm 0.22~\mathrm{GeV}$ for the $J^{PC} = 0^{++}$ state and $11.72 \pm 0.26~\mathrm{GeV}$ for the $J^{PC} = 1^{--}$ state.

 In the heavy-flavor sector, studies of the mass spectra of molecular-type hexaquark states have indicated the possible
existence of molecular configurations in the $DD^{*}$, $DD\bar{D}^{*}$, and $\Sigma_{c}\bar{D}^{(*)}$ systems, suggesting the emergence of numerous
deuteron-like hexaquarks~\cite{Wang:2024riu}. In Ref.~\cite{Geng:2024dpk}, the study of doubly charmed $H$-like dibaryon $\Lambda_{c}\Lambda_{c}$ scattering from lattice
QCD shows that the $\Lambda_{c}\Lambda_{c}$ system behaves as a typical scattering state with no bound state formation at the given
quark mass. Furthermore, the investigation of doubly-charmed hexaquark configurations was conducted within the
diquark framework, employing both the constituent quark model and the quark-interchange approach. The analysis,
based on key properties such as mass spectra, suggests the nonexistence of stable bound states. Nonetheless, the
evaluated root-mean-square radii indicate that these hexaquark systems exhibit a notably compact configuration~\cite{An:2025rjv}.
 The possibility of the existence of dibaryon states composed of heavy baryons, such as $\Xi^{*}_{c}\Xi^{*}_{c}$ and $\Xi_{c}\Xi_{c}$, has been
investigated~\cite{Lee:2011rka,Shiri:2023fjh,Shah:2024thr}. In Ref.~\cite{Shah:2024thr}, calculations based on a variational method yield masses of $5.291~\mathrm{GeV}$ for the
$J^{P} = 0^{+}$ channel with total isospin $I = 1$, and $5.281~\mathrm{GeV}$ for the $J^{P} = 1^{+}$ channel with $I = 0$.
Fully heavy dibaryons represent a particularly intriguing and debated topic. In Ref.~\cite{Richard:2020zxb}, the authors explored
the possibility of $bbbccc$ dibaryons and reported no bound states for either the $\Omega_{ccc}\Omega_{ccc}$ or $\Omega_{bbb}\Omega_{bbb}$ systems.
Conversely, Ref.~\cite{Lyu:2022tsd} indicated that the $\Omega_{ccc}\Omega_{ccc}$ system is weakly bound with a binding energy of $5.68(0.77)~\mathrm{MeV}$,
while Ref.~\cite{Mathur:2022ovu} predicted a deeply bound $\Omega_{bbb}\Omega_{bbb}$ state, with a binding energy of $81^{+14}_{-16}~\mathrm{MeV}$.
More recently, Ref.~\cite{Martin-Higueras:2024qaw} conducted a renewed analysis of the potential existence of $\Omega_{ccc}\Omega_{ccc}$ and $\Omega_{bbb}\Omega_{bbb}$
dibaryons using a constituent quark model. Their findings suggest that six charm or six bottom quarks can form bound states
with quantum numbers $J^{P} = 0^{+}$, yielding binding energies of $-0.71~\mathrm{MeV}$ for the charm case and $-1.98~\mathrm{MeV}$ for the bottom case.

 As outlined in the aforementioned studies, several theoretical frameworks have been utilized to investigate six-quark
systems and dibaryons. These include QCD sum rules, lattice QCD, chiral perturbation theory, quark models, and
effective field theories. Among these, the QCD sum rule approach has proven to be a powerful nonperturbative
method, extensively applied to various hadronic phenomena, including hadron spectroscopy and decay processes.
Within the QCD sum rule framework, suitable interpolating currents corresponding to the hadron of interest are
constructed. These currents are employed to formulate two-point or three-point correlation functions, which are used
to extract the hadron’s mass and decay properties, respectively. By matching the operator product expansion (OPE)
of the correlation function to its hadronic spectral representation, one derives sum rules that enable the determination
of physical observables.

One of the key advantages of the QCD sum rule method is that it is firmly grounded in the fundamental QCD
Lagrangian. The formalism is fully relativistic and free from arbitrary or purely phenomenological parameters, thereby
ensuring a high degree of theoretical consistency and predictive reliability.

In the light quark sector, the mass and coupling constant of a scalar six-quark state composed of $uuddss$ quarks
have been studied using the QCD sum rule approach to explore its viability as a dark matter candidate~\cite{Azizi:2019xla}.
The possibility of a $\Delta^{0}\Delta^{0}$ dibaryon has also been investigated through various methods, such as the quark
delocalization color screening model and the chiral quark model~\cite{Huang:2013nba}.
More recently, the scalar $\Delta^{0}\Delta^{0}$ state was studied using QCD sum rules~\cite{Mutuk:2022zgn}, where the mass was estimated as
$m_{D} = 2326^{+114}_{-126}~\mathrm{MeV}$, and the decay constant was found to be approximately
$f_{D} = \left( 2.94^{+0.30}_{-0.34} \right) \times 10^{-4}~\mathrm{GeV}^{7}$.
In the present work, we extend this line of investigation by calculating the mass and decay constant (residue)
of a hypothetical $\Delta^{0}\Delta^{0}$ dibaryon in the tensor state with quantum numbers $J^{P} = 2^{+}$ using the QCD sum rule method.

\section {METHODOLOGY}\label{sec:two}\label{sec:general} 
To investigate dibaryon systems within the framework of QCD sum rules, the $\Delta^{0}\Delta^{0}$ interpolating currents are constructed using the local Ioffe current representation of the $\Delta^{0}$ baryon:
\begin{equation}
\eta_\mu = \frac{1}{\sqrt{3}} \, \varepsilon^{abc} \left[ 2 \left( d_a^T C \gamma_\mu u_b \right) d_c + \left( d_a^T C \gamma_\mu d_b \right) u_c \right],
\end{equation} 
where $C = i\gamma_{2}\gamma_{0}$ is the charge conjugation matrix, 
$a$, $b$, $c$ denote color indices, $\gamma_{\mu}$ are Dirac gamma matrices, 
and $T$ indicates the transpose operator.
 
Using this interpolating current, the $\Delta^{0}\Delta^{0}$ dibaryon with quantum numbers $J^{P} = 2^{+}$ is constructed as
\begin{equation}
J_{\mu\nu}^{\Delta^0 \Delta^0} = \frac{1}{3} \, \epsilon^{abc} \epsilon^{def} \left[ 2 \left( d^{a\,T} C \gamma_\mu u^b \right) d^c + \left( d^{a\,T} C \gamma_\mu d^b \right) u^c \right] \cdot C \gamma_5 \cdot \left[ 2 d^{d\,T} \left( d^e C \gamma_\nu u^f \right) + u^{d\,T} \left( d^e C \gamma_\nu d^f \right) \right].
\end{equation}
To extract physical properties such as the mass and residue, 
we employ the two-point QCD sum rule method by analyzing the correlation function
\begin{equation}
\Pi_{\mu\nu\,\rho\sigma}(p^2) = i \int d^4x \, e^{i p \cdot x} \langle 0 | T \{ J^{\Delta^0\Delta^0}_{\mu\nu}(x) J^{\Delta^0\Delta^0 \dagger}_{\rho\sigma}(0) \} | 0 \rangle,
\end{equation}
where $T$ is the time-order operator and $J_{\mu\nu(\rho\sigma)}^{\Delta^{0}\Delta^{0}}$ 
is the interpolating current for the tensor dibaryon state. 
In the framework of QCD sum rules, the correlation function is evaluated from two complementary perspectives. 
On the physical (phenomenological) side, the correlation function is modeled in terms of hadronic degrees of freedom 
in the time-like region, allowing for the extraction of physical parameters such as the mass and residue in different states. 
On the QCD (theoretical) side, the same correlation function is computed in the space-like region using quark 
and gluon fields. This representation includes contributions from the QCD coupling constant, quark masses, and gluon 
condensates of various dimensions, capturing the non-perturbative aspects of the QCD vacuum. 
By equating the two sides through dispersion relations and invoking the quark-hadron duality assumption, 
one can extract the hadronic parameters in terms of QCD quantities. 
To enhance the reliability of the sum rule and suppress contributions from higher resonances and continuum states, 
Borel transformation and continuum subtraction procedures are employed. 
We commence by evaluating the hadronic (physical) representation of the correlation function through the insertion 
of appropriate complete sets of states, possessing the same quantum numbers as the interpolating current, into the 
corresponding positions. A complete set of intermediate hadronic states is inserted into the correlation function 
presented in Eq.~(3). After performing the integration over $x$, the correlation function can be written as:
\begin{equation}
\Pi^{\text{Phys}}_{\mu\nu\rho\sigma}(p) = \frac{\langle 0 | j_{\mu\nu}(0) | D(p) \rangle \langle D(p) | J_{\rho\sigma}(0) | 0 \rangle}{m_{D}^2 - p^2} + \cdots,
\end{equation}
where $\cdots$ represents the higher states and continuum contributions. 
The matrix elements responsible for creating hadronic states from the vacuum 
can be parameterized in terms of the decay constant and the corresponding 
polarization tensor as follows:
\begin{equation}
\langle 0 | J^{\Delta^0\Delta^0}_{\mu\nu} | D(p) \rangle = {f_{\Delta^0\Delta^0}} \varepsilon(p)^{\lambda}_{\mu\nu} + \cdots.
\end{equation}
Here, $f_{\Delta^{0}\Delta^{0}}$ denotes the coupling constant, and $\varepsilon_{\mu\nu}$ 
is the polarization tensor associated with the spin-2 state. 
The polarization sum is given by
\begin{equation}
\sum_{\lambda} \varepsilon^{(\lambda)}_{\mu\nu} \varepsilon^{*(\lambda)}_{\rho\sigma}
= \frac{1}{2} \left( T_{\mu\rho} T_{\nu\sigma} + T_{\mu\rho} T_{\nu\rho} \right) - \frac{1}{3} T_{\mu\nu} T_{\rho\sigma},
\end{equation}
where
\begin{equation}
T_{\mu\nu} = -g_{\mu\nu} + \frac{p_\mu p_\nu}{p^2}.
\end{equation}
Combining the above results and summing over polarization states via the corresponding completeness relation
yields the physical representation of the correlation function in the following final form
\begin{equation}
\Pi^{\text{Phys}}_{\mu\nu\rho\sigma}(p) = \frac{f_{\Delta^0\Delta^0}^2}{m_D^2 - p^2} \left[ \frac{1}{2} \left( g_{\mu\rho} g_{\nu\sigma} + g_{\mu\sigma} g_{\nu\rho} \right) \right] + \text{other structures} + \cdots,
\end{equation}
where the ellipses represent contributions from additional Lorentz structures, 
higher excited, and continuum states. 
The term proportional to $(g_{\mu\rho}g_{\nu\sigma} + g_{\mu\sigma}g_{\nu\rho})$ 
corresponds solely to the contribution from the spin-2 particle, 
while the remaining terms arise from the presence of spin-0 and spin-1 states as well.

 In the next step of our analysis, we evaluate the correlator $\Pi_{\mu\nu\rho\sigma}(p)$ 
using OPE. 
For this purpose, we substitute the explicit form of the current $J_{\mu\nu}(x)$ into Eq.~(3) 
and perform the necessary contractions of the quark fields via Wick’s theorem 
to derive $\Pi_{\mu\nu\rho\sigma}^{\mathrm{QCD}}(p)$. 
This leads to
\begin{equation}
\begin{aligned}
&\Pi^{\text{QCD}}_{\mu\nu\rho\sigma}(p) = \frac{16}{9} \int d^4x \, e^{ip \cdot x} \times \Big( \\
& \delta_{a f} \delta_{b e} \delta_{c d} \, \delta_{a' f'} \delta_{b' e'} \delta_{c' d'} \,
\operatorname{Tr}\big( S^{d}_{a d'}(x,0) \gamma^{\rho} \tilde{S}^{u}_{b e'}(x,0) \gamma^{\mu} \big) 
\operatorname{Tr}\big( S^{d}_{c c'}(x,0) \gamma^{5} \tilde{S}^{d}_{f f'}(x,0) \gamma^{5} \big) \\
& \qquad \times \operatorname{Tr}\big( S^{d}_{d a'}(x,0) \gamma^{\sigma} \tilde{S}^{u}_{e b'}(x,0) \gamma^{\nu} \big) \\
- & \delta_{a e} \delta_{b f} \delta_{c d} \, \delta_{a' f'} \delta_{b' e'} \delta_{c' d'} \,
\operatorname{Tr}\big( S^{d}_{a d'}(x,0) \gamma^{\rho} \tilde{S}^{u}_{b e'}(x,0) \gamma^{\mu} \big) 
\operatorname{Tr}\big( S^{d}_{c c'}(x,0) \gamma^{5} \tilde{S}^{d}_{f f'}(x,0) \gamma^{5} \big) \\
& \qquad \times \operatorname{Tr}\big( S^{d}_{d a'}(x,0) \gamma^{\sigma} \tilde{S}^{u}_{e b'}(x,0) \gamma^{\nu} \big) \\
- & \delta_{a f} \delta_{b d} \delta_{c e} \, \delta_{a' d'} \delta_{b' f'} \delta_{c' e'} \,
\operatorname{Tr}\big( S^{d}_{a d'}(x,0) \gamma^{\rho} \tilde{S}^{u}_{b e'}(x,0) \gamma^{\mu} \big) 
\operatorname{Tr}\big( S^{d}_{c c'}(x,0) \gamma^{5} \tilde{S}^{d}_{f f'}(x,0) \gamma^{5} \big) \\
& \qquad \times \operatorname{Tr}\big( S^{d}_{d a'}(x,0) \gamma^{\sigma} \tilde{S}^{u}_{e b'}(x,0) \gamma^{\nu} \big) \\
+ & \delta_{a d} \delta_{b f} \delta_{c e} \, \delta_{a' e'} \delta_{b' d'} \delta_{c' f'} \,
\operatorname{Tr}\big( S^{d}_{a d'}(x,0) \gamma^{\rho} \tilde{S}^{u}_{b e'}(x,0) \gamma^{\mu} \big) 
\operatorname{Tr}\big( S^{d}_{c c'}(x,0) \gamma^{5} \tilde{S}^{d}_{f f'}(x,0) \gamma^{5} \big) \\
& \qquad \times \operatorname{Tr}\big( S^{d}_{d a'}(x,0) \gamma^{\sigma} \tilde{S}^{u}_{e b'}(x,0) \gamma^{\nu} \big) \\
+ & \delta_{a e} \delta_{b d} \delta_{c f} \, \delta_{a' d'} \delta_{b' e'} \delta_{c' f'} \,
\operatorname{Tr}\big( S^{d}_{a d'}(x,0) \gamma^{\rho} \tilde{S}^{u}_{b e'}(x,0) \gamma^{\mu} \big) 
\operatorname{Tr}\big( S^{d}_{c c'}(x,0) \gamma^{5} \tilde{S}^{d}_{f f'}(x,0) \gamma^{5} \big) \\
& \qquad \times \operatorname{Tr}\big( S^{d}_{d a'}(x,0) \gamma^{\sigma} \tilde{S}^{u}_{e b'}(x,0) \gamma^{\nu} \big) \\
- & \delta_{a d} \delta_{b e} \delta_{c f} \, \delta_{a' d'} \delta_{b' e'} \delta_{c' f'} \,
\operatorname{Tr}\big( S^{d}_{a d'}(x,0) \gamma^{\rho} \tilde{S}^{u}_{b e'}(x,0) \gamma^{\mu} \big) 
\operatorname{Tr}\big( S^{d}_{c c'}(x,0) \gamma^{5} \tilde{S}^{d}_{f f'}(x,0) \gamma^{5} \big) \\
& \qquad \times \operatorname{Tr}\big( S^{d}_{d a'}(x,0) \gamma^{\sigma} \tilde{S}^{u}_{e b'}(x,0) \gamma^{\nu} \big) \\
- & \delta_{a f} \delta_{b e} \delta_{c d} \, \delta_{a' e'} \delta_{b' f'} \delta_{c' d'} \,
\operatorname{Tr}\big( S^{d}_{a d'}(x,0) \gamma^{\rho} \tilde{S}^{u}_{b e'}(x,0) \gamma^{\mu} \big) 
\operatorname{Tr}\big( S^{d}_{c c'}(x,0) \gamma^{5} \tilde{S}^{d}_{f f'}(x,0) \gamma^{5} \big) \\
& \qquad \times \operatorname{Tr}\big( S^{d}_{d a'}(x,0) \gamma^{\sigma} \tilde{S}^{u}_{e b'}(x,0) \gamma^{\nu} \big) \\
+ & \delta_{a e} \delta_{b f} \delta_{c d} \, \delta_{a' e'} \delta_{b' d'} \delta_{c' f'} \,
\operatorname{Tr}\big( S^{d}_{a d'}(x,0) \gamma^{\rho} \tilde{S}^{u}_{b e'}(x,0) \gamma^{\mu} \big) 
\operatorname{Tr}\big( S^{d}_{c c'}(x,0) \gamma^{5} \tilde{S}^{d}_{f f'}(x,0) \gamma^{5} \big) \\
& \qquad \times \operatorname{Tr}\big( S^{d}_{d a'}(x,0) \gamma^{\sigma} \tilde{S}^{u}_{e b'}(x,0) \gamma^{\nu} \big) \\
- & \delta_{a d} \delta_{b f} \delta_{c e} \, \delta_{a' f'} \delta_{b' e'} \delta_{c' d'} \,
\operatorname{Tr}\big( S^{d}_{a d'}(x,0) \gamma^{\rho} \tilde{S}^{u}_{b e'}(x,0) \gamma^{\mu} \big) 
\operatorname{Tr}\big( S^{d}_{c c'}(x,0) \gamma^{5} \tilde{S}^{d}_{f f'}(x,0) \gamma^{5} \big) \\
& \qquad \times \operatorname{Tr}\big( S^{d}_{d a'}(x,0) \gamma^{\sigma} \tilde{S}^{u}_{e b'}(x,0) \gamma^{\nu} \big) \\
+ & \delta_{a f} \delta_{b d} \delta_{c e} \, \delta_{a' f'} \delta_{b' d'} \delta_{c' e'} \,
\operatorname{Tr}\big( S^{d}_{a d'}(x,0) \gamma^{\rho} \tilde{S}^{u}_{b e'}(x,0) \gamma^{\mu} \big) 
\operatorname{Tr}\big( S^{d}_{c c'}(x,0) \gamma^{5} \tilde{S}^{d}_{f f'}(x,0) \gamma^{5} \big) \\
& \qquad \times \operatorname{Tr}\big( S^{d}_{d a'}(x,0) \gamma^{\sigma} \tilde{S}^{u}_{e b'}(x,0) \gamma^{\nu} \big) 
\Big) + \text{27638 similar terms}.
\end{aligned}
\end{equation}
Here, $\widetilde{S}(x) = C\, S^{T}(x)\, C$ and $S^{u(d)}(x)$ are the $u$ and $d$ quarks’ propagators, with color indices $a$, $b$, $\ldots$. The color deltas ensure the correct color contractions in the traces.
The next step in the QCD side involves employing the light-quark propagator in position space:
\begin{equation}
\begin{aligned}
S_q^{ab}(x) =\; & i \frac{\slashed{x}}{2\pi^2 x^4} \delta_{ab}
- \frac{m_q}{4\pi^2 x^2} \delta_{ab}
- \frac{\langle \bar{q} q \rangle}{12} \left(1 - i \frac{m_q}{4} \slashed{x} \right) \delta_{ab}
- \frac{x^2}{192} \langle \bar{q} g_s \sigma G q \rangle \left(1 - i \frac{m_q}{6} \slashed{x} \right) \delta_{ab} \\
& - \frac{i g_s G^{\mu\nu}_{ab}}{32 \pi^2 x^2} \left[ \slashed{x} \sigma_{\mu\nu} + \sigma_{\mu\nu} \slashed{x} \right] 
- \frac{\slashed{x} x^2 g_s^2 \langle \bar{q} q \rangle^2}{7776} \delta_{ab}
- \frac{x^4 \langle \bar{q} q \rangle \langle g_s^2 G^2 \rangle}{27648} \delta_{ab} \\
& + \frac{m_q g_s}{32 \pi^2} G^{\mu\nu}_{ab} \sigma_{\mu\nu} \left[ \ln\left( \frac{-x^2 \Lambda^2}{4} \right) + 2 \gamma_E \right] + \cdots.
\end{aligned}
\end{equation}
In these expressions, $q = u, d$, $\gamma_E \approx 0.577$ denotes the Euler–Mascheroni constant, 
$\Lambda$ is a scale parameter, 
$G^2 \equiv G^A_{\mu\nu} G^{\mu\nu}_A$, 
$G_{\mu\nu}^{ab} \equiv G^{\mu\nu}_A\, t^A_{ab}$, $A = 1, 2, \ldots, 8$, 
and $t^A = \lambda^A/2$, with $\lambda^A$ being the Gell–Mann matrices. 
The terms in Eq.~(10) include  the perturbative part, represented by the first term 
corresponding to the free propagator of a light quark, emission of a single gluon  and the non-perturbative contributions, 
which involve other gluonic effects such as the  two-gluon condensate. 
For the two–gluon condensate, $\langle 0| G^A_{\alpha\beta}(x) G^B_{\alpha'\beta'}(0) |0\rangle$, 
which leads to the four–dimension non–perturbative contribution, we consider the first term of the Taylor expansion 
for the gluon field at $x = 0$. We utilize~\cite{Barsbay:2022gtu}:
\begin{equation}
\langle 0 | G_{\alpha\beta}^A(0) \, G_{\alpha'\beta'}^{B}(0) | 0 \rangle 
= \frac{\langle G^2 \rangle}{96} \, \delta^{AB} \left( g_{\alpha\alpha'} g_{\beta\beta'} - g_{\alpha\beta'} g_{\alpha'\beta} \right),
\end{equation}
together with the SU(3) color identity:
\begin{equation}
t^{a}_{b} t^{a'}_{b'} = \frac{1}{2} (\delta_{a b'} \delta_{a' b} - \frac{1}{3} \delta_{a b} \delta_{a' b'}).
\end{equation}
In fact, the correlator $\Pi^{\mathrm{Phys}}(p^2)$ can be rewritten using the dispersion relation:
\begin{equation}
\Pi^{\text{Phys}}(p^2) = \int_{M'^2}^{\infty} \frac{\rho^{\text{Phys}}(s)}{s - p^2} ds + \cdots,
\end{equation}
with $M' = (4 m_d + 2 m_u)$. The spectral density is
\begin{equation}
\rho^{\text{Phys}}(s) = f_{\Delta^0\Delta^0}^2\, \delta(s - m_{\Delta^0\Delta^0}^2) + \rho^h(s)\, \theta(s - s_0),
\end{equation}
where $\theta(z)$ is the step function and $s_0$ is the continuum threshold. 
Higher resonances and continuum contributions to $\rho^{\mathrm{Phys}}(s)$ are represented through the unknown hadronic spectral density $\rho^h(s)$. It is evident that $\rho^{\mathrm{Phys}}(s)$ yields the following expression:
\begin{equation}
\Pi^{\text{Phys}}(p^2) = \frac{f_{\Delta^0\Delta^0}^2}{m_{\Delta^0\Delta^0} - p^2} + \int_{s_0}^{\infty} \frac{\rho^h(s)}{s - p^2} ds.
\end{equation}
The theoretical calculation of $\Pi^{\mathrm{OPE}}(p^2)$ is performed in the deep Euclidean region ($p^2 \ll 0$) 
via OPE. By analytically continuing $\Pi^{\mathrm{OPE}}(p^2)$ to the Minkowski domain 
and taking its imaginary part, the two–point spectral density $\rho^{\mathrm{OPE}}(s)$ is obtained, as presented below. 
In the deep Euclidean region ($p^2 \ll 0$), we perform the Borel transformation $\mathcal{B}$ 
to eliminate subtraction terms in the dispersion relation and to suppress the contributions from higher resonances 
and the continuum. Applying the Borel transform $\mathcal{B}$ in this domain yields:
\begin{equation}
\mathcal{B} \Pi^{\text{Phys}}(p^2) = f_{\Delta^0\Delta^0}^2\, e^{-m_{\Delta^0\Delta^0}^2/M^2} + \int_{s_0}^{\infty} \rho^h(s)\, e^{-s/M^2} \, ds,
\end{equation}
where $M^2$ denotes the Borel parameter and $s_0$ is the continuum threshold. 
To extract physical information from the correlation function, after applying the Borel transformation 
to both sides of the sum rule, the invariant amplitudes $\Pi^{\mathrm{Phys}}(p^2)$ and $\Pi^{\mathrm{QCD}}(p^2)$ are equated. 
Following this procedure, assuming hadron–parton duality, $\rho^h(s) \simeq \rho^{\mathrm{QCD}}(s)$ 
in the duality region, we subtract the continuum part from the QCD side, yielding:
\begin{equation}
f_{\Delta^0\Delta^0}^2 e^{-m_{\Delta^0\Delta^0}^2 / M^2} = \Pi^{\text{QCD}}(s_0, M^2),
\end{equation}
where
\begin{equation}
\Pi^{\text{QCD}}(s_0, M^2) = \int_{M'^2}^{s_0} ds \,\rho(s) \, \ e^{-s/M^2}.
\end{equation}
The spectral density $\rho(s)$ corresponds to the imaginary part of $\Pi^{\mathrm{OPE}}(p^2)$ 
and is given by $\rho(s) = \frac{1}{\pi} \mathrm{Im}[\Pi(s)]$. 
Under the assumptions $m_u \to 0$ and $m_d \to 0$, the explicit form of $\rho(s)$ reads:
\begin{align}
\rho(s) = & - \frac{47\, s^{7}}{6242697216000\,\pi^{10}} + \frac{527\,\langle G^{2}\rangle\, s^{5}\, g_s^{2}}{1651507200\,\pi^{10}} + \frac{s^{4}\,\langle\bar q q\rangle^{2}\, g_s^{2}}{376233984\,\pi^{8}} - \frac{1961\, s^{4}\,\langle\bar q q\rangle^{2}}{209018880\,\pi^{6}} - \frac{331\, s^{4}\,\langle\bar q q\rangle^{2}}{52254720\,\pi^{6}} \nonumber \\
& - \frac{41\, s^{4}\,\langle\bar q q\rangle^{2}\, g_s^{2}}{6449725440\,\pi^{8}} - \frac{23\, s^{4}\,\langle\bar q q\rangle^{2}}{20901888\,\pi^{6}} + \frac{679\, s^{3}\, \langle\bar q q\rangle^{2}\, m_0^{2}}{7464960\,\pi^{6}} + \frac{13\, s^{3}\, \langle\bar q q\rangle^{2}\, m_0^{2}}{207360\,\pi^{6}} + \frac{17\, s^{3}\, m_0^{2}\, \langle\bar q q\rangle^{2}}{1492992\,\pi^{6}}.
\end{align}
In the following, using the shorthand
\begin{equation}
\Pi'(s_0, M^2) = \frac{d}{d(-1/M^2)} \, \Pi(s_0, M^2),
\end{equation}
and after straightforward algebra, we obtain the mass and the residue
\begin{equation}
m^2_{\Delta^0 \Delta^0} =
\frac{\Pi'(s_0, M^2)}{\Pi(s_0, M^2)} =
\frac{\displaystyle \int_{M'^2}^{s_0} ds \, s \, \rho(s) \, e^{-s/M^2}}
{\displaystyle \int_{M'^2}^{s_0} ds \, \rho(s) \, e^{-s/M^2}},
\end{equation}
and
\begin{equation}
f_{\Delta^0 \Delta^0}^2 = e^{m_{\Delta^0 \Delta^0}^2 / M^2} \, \Pi(s_0, M^2).
\end{equation}
\section{Numerical analysis}\label{sec:numerical}
For the numerical evaluation of the obtained expressions for the mass and residue of the $\Delta^{0}\Delta^{0}$ 
dibaryon states with $J^{P} = 2^{+}$, a set of input parameters, including the gluon and quark condensates, 
is required. These parameters are listed in Table~\text{I}.
\begin{table}[!htb]
\centering	
		\begin{tabular}{c*{12}{c}r}	
			\hline
			\hline
			$\langle \bar{q} q \rangle$ 
			& \qquad $-(0.24 \pm 0.01)^3~\text{GeV}^3$ \cite{Belyaev:1982sa}
			& \qquad $m_0^2$ 
			& \qquad
			$(0.8 \pm 0.1)~\text{GeV}^2$\cite{Belyaev:1982sa} 
			& \qquad $\langle \alpha_s G^2 \rangle$
			& \qquad $(0.012 \pm 0.004)~\text{GeV}^4$ \cite{Belyaev:1982cd}
			\\
			$\langle \alpha_s \rangle$
			& \qquad $(0.118\pm 0.005)$ \cite{Furstenau:1994ra} 
			& \qquad $\langle g_s^2 \rangle$ 
			& \qquad $4\pi \alpha_s$
			\\
			\hline
\end{tabular}
\caption{Key input parameters and their values used in the analysis.}
\label{table:inputPar}
\end{table}

 At this point in the analysis, it is crucial to determine the appropriate working windows for the two auxiliary
parameters involved in QCD sum rule calculations of physical observables, namely the Borel parameter $M^2$ and the
continuum threshold $s_0$. While these parameters are not physical in themselves, they are subject to specific constraints
within the QCD sum rule framework and cannot be selected arbitrarily.

A key requirement is that physical quantities such as masses and residue should remain largely stable within certain
ranges of $M^2$ and $s_0$. Identifying these stability regions is essential for ensuring the reliability of the results, as it
helps minimize theoretical uncertainties and validates the consistency of the QCD sum rule approach.

The parameters must be selected to ensure that the pole contribution (PC) dominates in the physical observables
extracted via the sum rule method. Additionally, the stability of these results with respect to $M^2$ and the convergence
of the OPE are critical constraints in the numerical analysis. These conditions can be quantitatively assessed by
introducing the following expressions:
\begin{equation}
\text{PC} = \frac{\Pi(M^2, s_0)}{\Pi(M^2, \infty)},
\end{equation}
and
\begin{equation}
R(M^2) = \frac{\Pi^{\text{Dim}N}(M^2, s_0)}{\Pi(M^2, s_0)}.
\end{equation}
Here, we impose the following criteria: The upper bound of the Borel parameter $M^2$ is determined by ensuring
that the pole contribution is considerably large compared to each of the higher excited states and continuum. In technical
language, we require that the pole contribution be as maximum as possible in the exotic dibaryon channel: We set $\mathrm{PC} \geq 0.2$. Moreover, the lower bound of $M^2$ is determined by
the requirement of convergence of the OPE. This criterion implies that the perturbative
contribution must dominate over the non-perturbative terms, and the effects of higher-dimensional operators should
remain suppressed. To ensure this, we require that the combined contribution of the last three non-perturbative
terms does not exceed $5\%$ of the total (perturbative plus non-perturbative) contributions, i.e., $R(M^2) \leq 0.05$. Moreover,
 the chosen range for $s_0$ should lie significantly above the production threshold of two $\Delta^{0}$ baryons, which is
approximately $4 m_{\Delta^{0}}^2 \simeq 6.07~\mathrm{GeV}^2$. The range of $s_0$ may be selected as the minimal value that ensures a stable and
reliable Borel window. Alternatively, $s_0$ can be fixed by fulfilling additional constraints so as to maximize the PC. Note that for multiquark hadrons, or more precisely exotic states, the available information is
often scarce. In such situations, $s_0$ can be estimated based on the aforementioned constraints.

 We then choose the parameters $(2 m_{\Delta^{0} \Delta^{0}} + 0.2)^2~\mathrm{GeV}^2 \leq s_0 \leq (2 m_{\Delta^{0} \Delta^{0}} + 0.4)^2~\mathrm{GeV}^2$, 
which define an energetically feasible window for exciting the $\Delta^{0} \Delta^{0}$ system to its first excited state. 
In addition, the Borel parameter is chosen as $8.0~\mathrm{GeV}^2 \leq M_B^2 \leq 10.0~\mathrm{GeV}^2$, 
which ensures the necessary conditions for a reliable QCD sum rule analysis.

Figures~1 and~2 present the mass and decay constant of the $\Delta^{0} \Delta^{0}$ dibaryon as functions of $M^2$ and $s_0$. 
Figure~1 (left panel) demonstrates that the mass remains relatively stable across the chosen range of the 
Borel parameter $M^2$ and shows very good stability. 
In the right panel of Figure~1, a mild dependence of the mass on the continuum threshold $s_0$ is observed. 
As shown in Figure~2, though  the residue shows a good stability with respect to the variations  of  $M^2$ in its working region,  it indicates a residual sensitivity  to variations in $s_0$. As a result, the requirements of the method are  satisfied. 
\begin{figure}[H]
    \centering
    \includegraphics[width=0.45\textwidth]{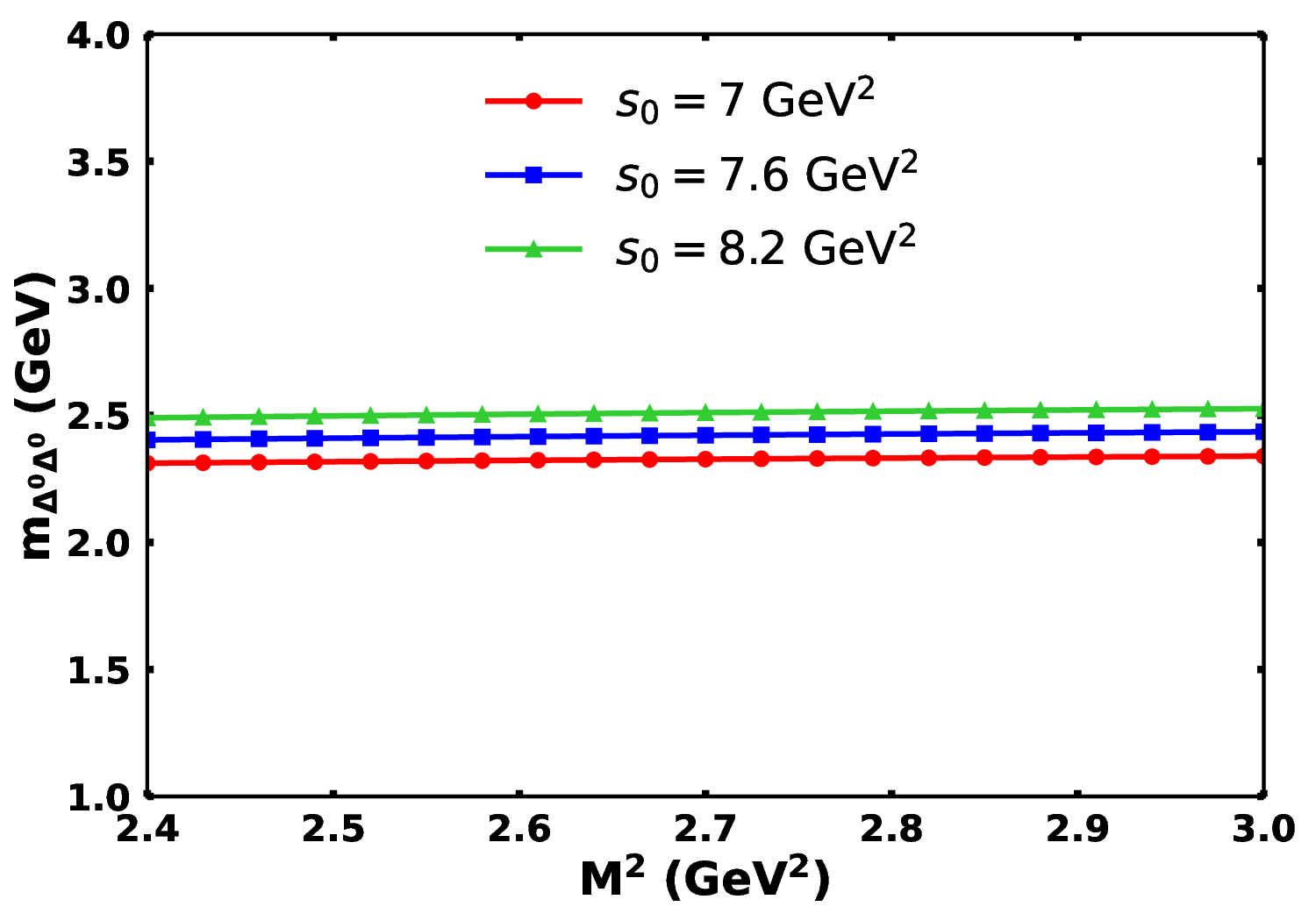}
    \hspace{0.02\textwidth} 
    \includegraphics[width=0.45\textwidth]{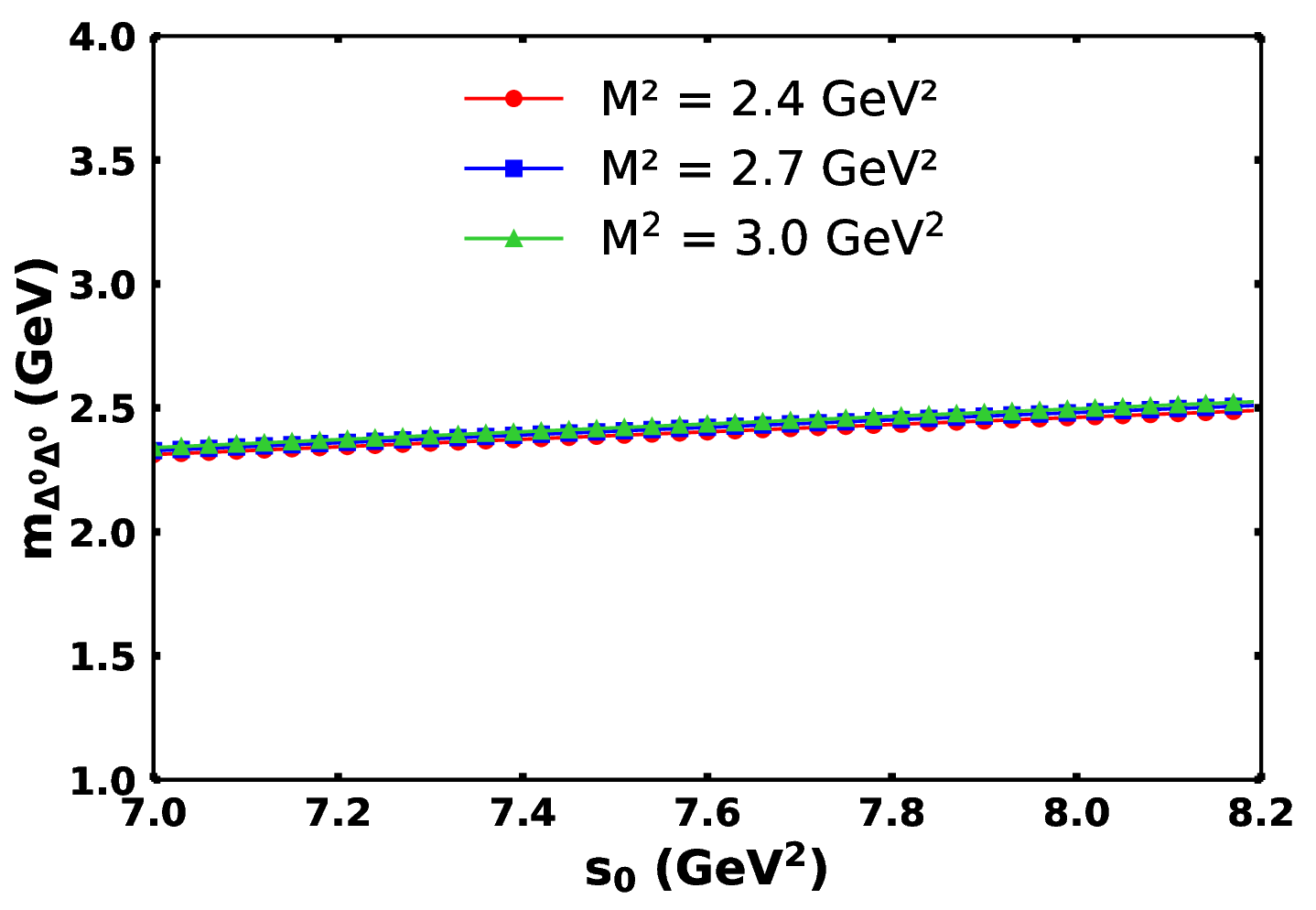}
    \caption{The left panel shows the variation of the $\Delta^0\Delta^0$ dibaryon mass with respect to $M^2$ at fixed $s_0$, while the right panel illustrates its dependence on $s_0$ at fixed $M^2$.}
    \label{fig:dibaryon_mass_combined1}
\end{figure}
\begin{figure}[H]
    \centering
    \includegraphics[width=0.45\textwidth]{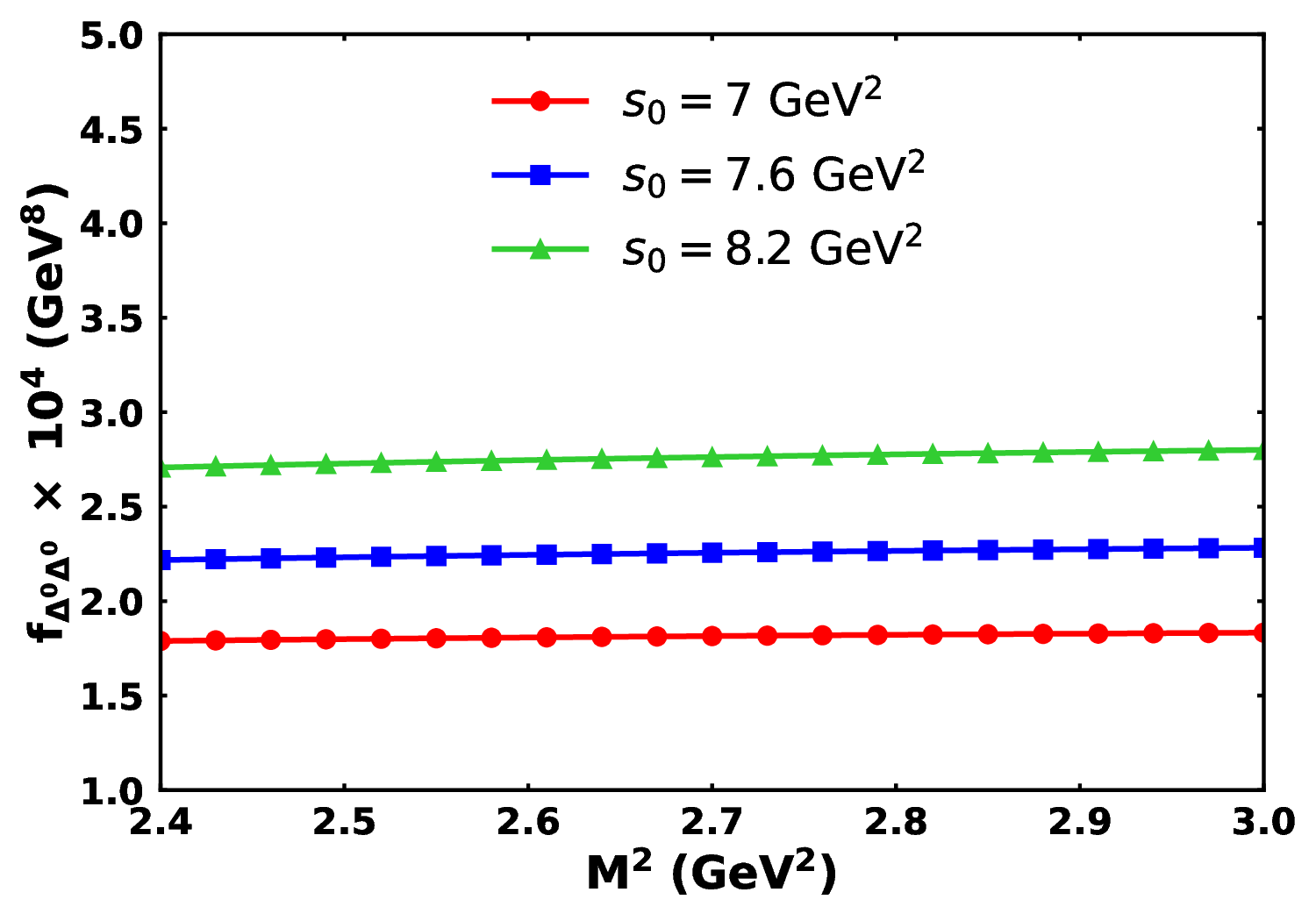}
    \hspace{0.02\textwidth} 
    \includegraphics[width=0.45\textwidth]{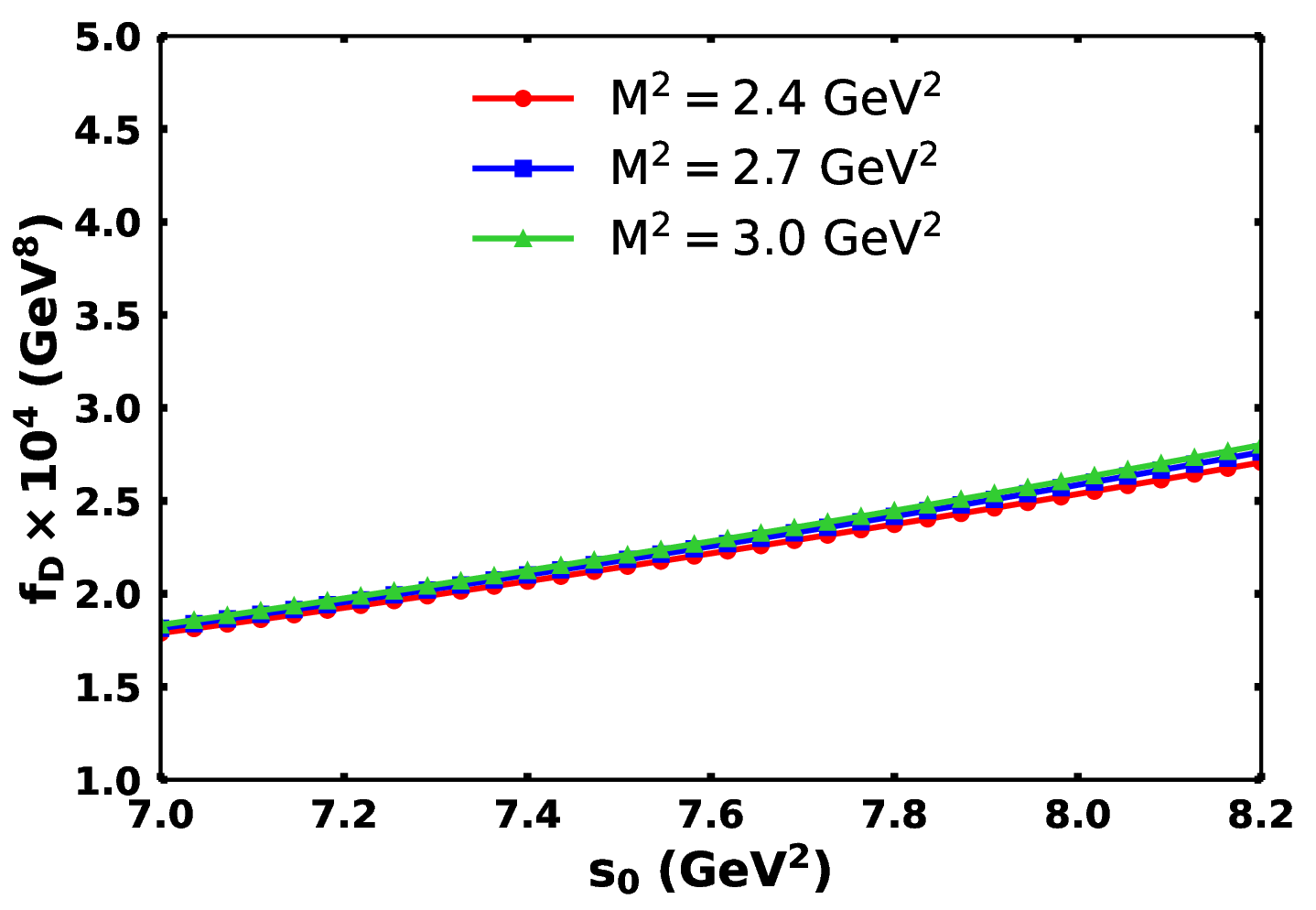}
    \caption{The left panel shows the variation of the $\Delta^0\Delta^0$ dibaryon mass with respect to $M^2$ at fixed $s_0$, while the right panel illustrates its dependence on $s_0$ at fixed $M^2$.}
    \label{fig:dibaryon_mass_combined1}
\end{figure}

To extract the mass $m_{\Delta^{0} \Delta^{0}}$ and residue $f_{\Delta^{0} \Delta^{0}}$ of the $\Delta^{0} \Delta^{0}$ dibaryon with 
$J^{P} = 2^{+}$, we perform calculations using the range of values for the Borel parameter $M^2$ and the continuum 
threshold $s_0$. The final values are obtained by averaging over the working intervals defined above. 
The results are as follows:
\begin{equation}
m_{\Delta^0\Delta^0} = 2426^{+101}_{-108}~\text{MeV}, \quad
f_{\Delta^0\Delta^0} = \left(2.30^{+0.49}_{-0.45}\right) \times 10^{-4}~\text{GeV}^8
\end{equation}
The uncertainties in the mass and decay residue arise from the variation between the extreme values of 
$m_{\Delta^{0} \Delta^{0}}$ and $f_{\Delta^{0} \Delta^{0}}$. 
The central mass value lies approximately $38~\mathrm{MeV}$ below the $2 m_{\Delta^{0}}$ threshold, 
indicating a possible bound molecular state of the scalar $\Delta^{0} \Delta^{0}$ dibaryon. 
Compared to the calculated $\Delta^{0} \Delta^{0}$ dibaryon in the $J^{P} = 0^{+}$ channel, 
where a mass of $m_{\Delta^{0} \Delta^{0}} = 2326^{+114}_{-126}~\mathrm{MeV}$ has been reported~\cite{Mutuk:2022zgn}, 
the mass obtained in the $J^{P} = 2^{+}$ channel is higher by about $100~\mathrm{MeV}$, 
leading to a weaker binding. 
It is also worth noting that the $\Omega \Omega$ dibaryon, an exotic particle composed entirely of strange quarks, 
has been investigated in both the $J^{P} = 0^{+}$ and $J^{P} = 2^{+}$ channels~\cite{Chen:2019vdh}. 
In contrast to the $\Delta^{0} \Delta^{0}$ dibaryon, the predicted mass of the $\Omega \Omega$ state 
in the tensor channel is lower than that in the scalar channel, 
indicating a stronger binding for the tensor state.

It is important to emphasize that the present analysis excludes the influence of radiative corrections. 
Although the mass, extracted from the ratio of two sum rules, is relatively insensitive to such corrections, 
the residue calculated from an individual sum rule can exhibit considerable fluctuations. 
Furthermore, next-to-leading order (NLO) perturbative corrections to baryon sum rules have been reported 
as significant~\cite{Krasnikov:1982ea,Groote:1999zp}. 
Similarly, while NLO $\alpha_s$ corrections to the perturbative component in the OPE
for light tetraquark currents are sizable, their overall impact on the sum rules remains limited owing to the 
predominance of nonperturbative condensate contributions~\cite{Groote:2014pva}. 
Furthermore, given that the analysis of the $\Delta^{0} \Delta^{0}$ dibaryon with $J^{P} = 2^{+}$ suggests the existence 
of a bound state, the estimation of its size can be interesting. 
Based on the definition of binding energy, the size of the $\Delta^{0} \Delta^{0}$ dibaryon can be determined 
using the following expression:
\begin{equation}
r \sim \frac{1}{\sqrt{2 \mu \, B_E}}
\end{equation}
where the binding energy is $B_E = 2m_{\Delta^0} - m_{\Delta^0\Delta^0}$.
In this expression, $r$ represents the distance between the components, while 
$\mu = \frac{m_1 m_2}{m_1 + m_2}$ corresponds to the reduced mass of the two-hadron system. 
The size of the tensor state is found to be approximately $r \simeq 0.90~\mathrm{fm}$, 
nearly twice that of the $\Delta^0 \Delta^0$ dibaryon in the $J^P = 0^+$ state. As is clear, both the tensor and scalar states remain below the confinement radius of 1 fm ~\cite{Guo:2017jvc}. In the case of the tensor state, while it may suggest a molecular configuration, 
the radius indicates an overlap between two $\Delta^{0}$ baryons, making the molecular 
description questionable. The obtained size suggests a compact hexaquark 
structure for this particle. However, calculating other properties of this state 
may help in achieving a better understanding of its internal configuration. In contrast to the $\Delta^{0} \Delta^{0}$ dibaryon, which represents a light-quark exotic dibaryon, it is of particular interest 
to estimate the radius of the $\Omega \Omega$ dibaryon mentioned above. 
Based on the binding energy calculated in Ref.~\cite{Chen:2019vdh}, and assuming the same quantum state as considered in our study, 
the resulting radii are found to be approximately $r \simeq 1.24~\mathrm{fm}$ for the $J^{P} = 0^{+}$ state 
and $r \simeq 0.97~\mathrm{fm}$ for the $J^{P} = 2^{+}$ state. 
These results indicate that the tensor state remains within the typical confinement radius.

\section{CONCLUDING NOTES}\label{sec:con}
The investigation of newly observed exotic hadronic states constitutes a primary focus within contemporary hadron 
physics research. In this study, we explore tensor $\Delta^{0} \Delta^{0}$ dibaryon states with spin-parity $J^{P} = 2^{+}$ 
using the framework of QCD sum rules. We construct interpolating currents for the tensor $\Delta^{0} \Delta^{0}$ dibaryon, 
and calculate the corresponding spectral densities and two-point correlation functions, including contributions from 
condensates up to dimension nine. The numerical analysis indicates that the tensor current leads to stable mass sum rules. Based on this, we provide a 
reliable mass prediction for the tensor $\Delta^{0} \Delta^{0}$ dibaryon with $J^{P} = 2^{+}$, obtaining
\(
m_{\Delta^{0} \Delta^{0},\, 2^{+}} =  2426^{+101}_{-108}~\mathrm{MeV}.
\)
This result points to the possible existence of a bound state, with an estimated binding energy of approximately 
$38~\mathrm{MeV}$. The residue was determined to be
\(
f_{\Delta^{0} \Delta^{0}} = \left(2.30^{+0.49}_{-0.45}\right) \times 10^{-4}~\mathrm{GeV}^8.
\)
Additionally, we estimated the spatial size of the tensor $\Delta^{0} \Delta^{0}$ dibaryon to be approximately 
$r \simeq 0.90~\mathrm{fm}$. The calculated binding energy suggests that the tensor $\Delta^{0} \Delta^{0}$ dibaryon is a 
potentially stable system. We hope that these findings will contribute to future theoretical investigations and 
experimental searches for dibaryon states.

	

\begin{thebibliography}{99}
		
		
		
\bibitem{Gell-Mann:1964ewy}
M.~Gell-Mann,
``A Schematic Model of Baryons and Mesons,''
\href{https://doi.org/10.1016/S0031-9163(64)92001-3}{Phys. Lett. \textbf{8}, 214-215 (1964)},
		
		
\bibitem{Belle:2003nnu}
S.~K.~Choi \textit{et al.} [Belle],
``Observation of a narrow charmonium-like state in exclusive $B^\pm \to K^\pm \pi^+ \pi^- J/\psi$ decays,''
\href{https://10.1103/PhysRevLett.91.262001}{Phys. Rev. Lett. \textbf{91}, 262001 (2003)},
\href{https://arxiv.org/pdf/hep-ex/0309032}{[arXiv:hep-ex/0309032 [hep-ex]].}
		
		
\bibitem{Swanson:2006st}
E.~S.~Swanson,
``The New heavy mesons: A Status report,''
\href{https://10.1016/j.physrep.2006.04.003}{Phys. Rept. \textbf{429}, 243-305 (2006)}
\href{https://arxiv.org/pdf/hep-ph/0601110}{[arXiv:hep-ph/0601110 [hep-ph]].}
		
		
		
		
\bibitem{LHCb:2015yax}
R.~Aaij \textit{et al.} [LHCb],
``Observation of $J/\psi p$ Resonances Consistent with Pentaquark States in $\Lambda_b^0 \to J/\psi K^- p$ Decays,''
\href{https://10.1103/PhysRevLett.115.072001}{Phys. Rev. Lett. \textbf{115}, 072001 (2015)}
\href{https://arxiv.org/pdf/1507.03414}{[arXiv:1507.03414 [hep-ex]].}
		
\bibitem{Chen:2016qju}
H.~X.~Chen, W.~Chen, X.~Liu and S.~L.~Zhu,
\href{https://10.1016/j.physrep.2016.05.004}{Phys. Rept. \textbf{639}, 1-121 (2016)}
\href{https://arxiv.org/pdf/1601.02092}{[arXiv:1601.02092 [hep-ph]].}
		
\bibitem{Lebed:2016hpi}
R.~F.~Lebed, R.~E.~Mitchell and E.~S.~Swanson,
``Heavy-Quark QCD Exotica,''
\href{https://doi:10.1016/j.ppnp.2016.11.003}{Prog. Part. Nucl. Phys. \textbf{93}, 143-194 (2017)}
\href{https://arxiv.org/pdf/1610.04528}{[arXiv:1610.04528 [hep-ph]].}
		
\bibitem{Olsen:2017bmm}
S.~L.~Olsen, T.~Skwarnicki and D.~Zieminska,
``Nonstandard heavy mesons and baryons: Experimental evidence,''
\href{https://10.1103/RevModPhys.90.015003}{Rev. Mod. Phys. \textbf{90}, no.1, 015003 (2018)}
\href{https://arxiv.org/pdf/1708.04012}{[arXiv:1708.04012 [hep-ph]].}
		
\bibitem{Urey:1932gik}
H.~C.~Urey, F.~G.~Brickwedde and G.~M.~Murphy,
``A Hydrogen Isotope of Mass 2,''
\href{https://doi.org/10.1103/physrev.39.164}{Phys. Rev. \textbf{39}, no.1, 164-165 (1932)},

\bibitem{Jaffe:1976yi}
R.~L.~Jaffe,
``Perhaps a Stable Dihyperon,''
\href{https://10.1103/PhysRevLett.38.195}{Phys. Rev. Lett. \textbf{38}, 195-198 (1977)}
								

\bibitem{WASA-at-COSY:2011bjg}
P.~Adlarson \textit{et al.} [WASA-at-COSY],
``ABC Effect in Basic Double-Pionic Fusion --- Observation of a new resonance?,''
\href{https://10.1103/PhysRevLett.106.242302}{Phys. Rev. Lett. \textbf{106}, 242302 (2011)},
\href{https://arxiv.org/pdf/1104.0123}{[arXiv:1104.0123 [nucl-ex]].}
		
\bibitem{WASA-at-COSY:2012seb}
P.~Adlarson \textit{et al.} [WASA-at-COSY],
``Isospin Decomposition of the Basic Double-Pionic Fusion in the Region of the ABC Effect,''
\href{https://10.1016/j.physletb.2013.03.019}{Phys. Lett. B \textbf{721}, 229-236 (2013)},
\href{https://arxiv.org/pdf/1212.2881}{[arXiv:1212.2881 [nucl-ex]].}
		
\bibitem{Gal:2013dca}
A.~Gal and H.~Garcilazo,
``Three-Body Calculation of the Delta-Delta Dibaryon Candidate D(03) at 2.37 GeV,''
\href{https://10.1103/PhysRevLett.111.172301}{Phys. Rev. Lett. \textbf{111}, 172301 (2013)},
\href{https://arxiv.org/pdf/1308.2112}{[arXiv:1308.2112 [nucl-th]].}

\bibitem{WASA-at-COSY:2014dmv}
P.~Adlarson \textit{et al.} [WASA-at-COSY],
``Evidence for a New Resonance from Polarized Neutron-Proton Scattering,''
\href{https://10.1103/PhysRevLett.112.202301}{Phys. Rev. Lett. \textbf{112}, no.20, 202301 (2014)},
\href{https://arxiv.org/pdf/1402.6844}{[arXiv:1402.6844 [nucl-ex]].}

\bibitem{WASA-at-COSY:2014qkg}
P.~Adlarson \textit{et al.} [WASA-at-COSY],
``Measurement of the $np \to np\pi^0\pi^0$ Reaction in Search for the Recently Observed $d^*(2380)$ Resonance,''
\href{https://doi.org/10.1016/j.physletb.2015.02.067}{Phys. Lett. B \textbf{743}, 325-332 (2015)},
\href{https://arxiv.org/pdf/1409.2659}{[arXiv:1409.2659 [nucl-ex]].}
	

		
\bibitem{Dong:2015cxa}
Y.~Dong, P.~Shen, F.~Huang and Z.~Zhang,
``Theoretical study of the $d^*(2380) \to d \pi \pi$ decay width,''
\href{https://doi.org/10.1103/PhysRevC.91.064002}{Phys. Rev. C \textbf{91}, no.6, 064002 (2015)},
\href{https://arxiv.org/pdf/1503.02456}{[arXiv:1503.02456 [nucl-th]].}


\bibitem{Dong:2018ryf}
Y.~Dong, P.~Shen, F.~Huang and Z.~Zhang,
``Study of d\ensuremath{*}(2380) resonance in a chiral constituent quark model,''
\href{https://doi.org/10.1142/S0217751X18300314}{Int. J. Mod. Phys. A \textbf{33}, no.33, 1830031 (2018)},


\bibitem{Dong:2019zvg}
Y.~Dong, P.~Shen, F.~Huang and Z.~Zhang,
``The Properties of d*(2380) in a Hexaquark Scenario,''
\href{https://doi.org/10.7566/JPSCP.26.022016}{JPS Conf. Proc. \textbf{26}, 022016 (2019)},


\bibitem{Chen:2014vha}
H.~X.~Chen, E.~L.~Cui, W.~Chen, T.~G.~Steele and S.~L.~Zhu,
``QCD sum rule study of the d*(2380),''
\href{https://doi.org/10.1103/PhysRevC.91.025204}{}Phys. Rev. C \textbf{91}, no.2, 025204 (2015)
doi:10.1103/PhysRevC.91.025204
\href{https://arxiv.org/pdf/1410.0394}{[arXiv:1410.0394 [hep-ph]].}
	
\bibitem{Huang:2019lzt}
F.~Huang, Y.~B.~Dong, P.~N.~Shen and Z.~Y.~Zhang,
``Understanding d*(2380) in a chiral quark model,''
\href{https://doi.org/10.1051/epjconf/201919902017}{EPJ Web Conf. \textbf{199}, 02017 (2019)}

\bibitem{Bashkanov:2020lxv}
M.~Bashkanov, G.~Clash, M.~Mocanu, M.~Nicol and D.~P.~Watts,
``Decay properties of the $d^*(2380)$ hexaquark multiplet,''
\href{https://arxiv.org/pdf/2012.11449}{[arXiv:2012.11449 [hep-ph]].}
	

	
\bibitem{Kim:2020rwn}
H.~Kim, K.~S.~Kim and M.~Oka,
``Hexaquark picture for $d^*(2380)$,''
\href{https://doi.org/10.1103/PhysRevD.102.074023}{Phys. Rev. D \textbf{102}, no.7, 074023 (2020)},
\href{https://arxiv.org/pdf/2009.11983}{[arXiv:2009.11983 [hep-ph]].}

\bibitem{Celi:2023gtj}
M.~O.~Celi, M.~Bashkanov, M.~Mariani, M.~G.~Orsaria, A.~Pastore, I.~F.~Ranea-Sandoval and F.~Weber,
``Destabilization of high-mass neutron stars by the emergence of d*-hexaquarks,''
\href{https://doi.org/10.1103/PhysRevD.109.023004}{Phys. Rev. D \textbf{109}, no.2, 023004 (2024)},
\href{https://arxiv.org/pdf/2312.03880}{[arXiv:2312.03880 [nucl-th]].}


\bibitem{Celi:2025wnc}
M.~O.~Celi, M.~Mariani, R.~Kumar, M.~Bashkanov, M.~G.~Orsaria, A.~Pastore, I.~F.~Ranea-Sandoval and V.~Dexheimer,
\href{https://doi.org/10.1103/3lyv-45jp}{``Exploring the role of d* hexaquarks on quark deconfinement and hybrid stars,''
Phys. Rev. D \textbf{112}, no.2, 023027 (2025)},
\href{https://arxiv.org/pdf/2504.00981}{[arXiv:2504.00981 [nucl-th]].}

\bibitem{Shahrbaf:2022upc}
M.~Shahrbaf, D.~Blaschke, S.~Typel, G.~R.~Farrar and D.~E.~Alvarez-Castillo,
``Sexaquark dilemma in neutron stars and its solution by quark deconfinement,''
\href{https://doi.org/10.1103/PhysRevD.105.103005}{Phys. Rev. D \textbf{105}, no.10, 103005 (2022)},
\href{https://arxiv.org/pdf/2202.00652}{[arXiv:2202.00652 [nucl-th]].}

	
\bibitem{Farrar:2003qy}
G.~R.~Farrar and G.~Zaharijas,
``Nuclear and nucleon transitions of the H dibaryon,''
\href{https://doi.org/10.1103/PhysRevD.70.014008}{Phys. Rev. D \textbf{70}, 014008 (2004)},
\href{https://arxiv.org/pdf/hep-ph/0308137}{[arXiv:hep-ph/0308137 [hep-ph]].}
				

\bibitem{Farrar:2017eqq}
G.~R.~Farrar,
``Stable Sexaquark,''
\href{https://arxiv.org/pdf/1708.08951}{[arXiv:1708.08951 [hep-ph]].}
		
		
\bibitem{Farrar:2018hac}
G.~R.~Farrar,
``A precision test of the nature of Dark Matter and a probe of the QCD phase transition,''
\href{https://arxiv.org/pdf/1805.03723}{[arXiv:1805.03723 [hep-ph]].}
		
		
\bibitem{Halprin:1982pb}
A.~Halprin and A.~k.~Kerman,
``LEPTON BAGS,''
\href{https://doi.org/10.1103/PhysRevD.26.2532}{Phys. Rev. D \textbf{26}, 2532-2533 (1982)}
		
\bibitem{Yost:1985mj}
S.~A.~Yost and C.~R.~Nappi,
``The Mass of the $H$ Dibaryon in a Chiral Model,''
\href{https://doi.org/10.1103/PhysRevD.32.816}{Phys. Rev. D \textbf{32}, 816 (1985)}
		
\bibitem{Oka:1986fr}
M.~Oka, K.~Shimizu and K.~Yazaki,
``Hyperon - Nucleon and Hyperon-hyperon Interaction in a Quark Model,''
\href{https://doi.org/10.1016/0375-9474(87)90371-X}{Nucl. Phys. A \textbf{464}, 700-716 (1987)}
doi:10.1016/0375-9474(87)90371-X
		
\bibitem{Straub:1988mz}
U.~Straub, Z.~Y.~Zhang, K.~Brauer, A.~Faessler and S.~B.~Khadkikar,
``Binding Energy of the Dihyperon in the Quark Cluster Model,''
\href{https://doi.org/10.1016/0370-2693(88)90763-0}{Phys. Lett. B \textbf{200}, 241-245 (1988)}
		
\bibitem{Koike:1989ak}
Y.~Koike, K.~Shimizu and K.~Yazaki,
``Study of Hyperon - Nucleon and Hyperon-hyperon Interaction in the Flipflop Model,''
\href{https://doi.org/10.1016/0375-9474(90)90403-9}{Nucl. Phys. A \textbf{513}, 653-666 (1990)}
doi:10.1016/0375-9474(90)90403-9
		
\bibitem{Kodama:1994np}
N.~Kodama, M.~Oka and T.~Hatsuda,
``H dibaryon in the QCD sum rule,''
\href{https://doi.org/10.1016/0375-9474(94)90908-3}{Nucl. Phys. A \textbf{580}, 445-454 (1994)}
\href{https://arxiv.org/pdf/hep-ph/9404221}{[arXiv:hep-ph/9404221 [hep-ph]].}
		
		
\bibitem{Khrykin:2000gh}
A.~S.~Khrykin, V.~F.~Boreiko, Y.~G.~Budyashov, S.~B.~Gerasimov, N.~V.~Khomutov, Y.~G.~Sobolev and V.~P.~Zorin,
``Search for N N decoupled dibaryons using the process p p ---\ensuremath{>} gamma gamma X below the pion production threshold,''
\href{https://doi.org/10.1103/PhysRevC.64.034002}{Phys. Rev. C \textbf{64}, 034002 (2001)},
\href{https://arxiv.org/pdf/nucl-ex/0012011}{[arXiv:nucl-ex/0012011 [nucl-ex]].}
		
\bibitem{Bashkanov:2008ih}
M.~Bashkanov, C.~Bargholtz, M.~Berlowski, D.~Bogoslawsky, H.~Calen, H.~Clement, L.~Demiroers, E.~Doroshkevich, D.~Duniec and C.~Ekstrom, \textit{et al.}
``Double-Pionic Fusion of Nuclear Systems and the ABC Effect: Aproaching a Puzzle by Exclusive and Kinematically Complete Measurements,''
\href{https://doi.org/10.1103/PhysRevLett.102.052301}{Phys. Rev. Lett. \textbf{102}, 052301 (2009)},
doi:10.1103/PhysRevLett.102.052301
\href{https://arxiv.org/pdf/0806.4942}{[arXiv:0806.4942 [nucl-ex]].}
	
\bibitem{Vijande:2011im}
J.~Vijande, A.~Valcarce and J.~M.~Richard,
``Stability of hexaquarks in the string limit of confinement,''
\href{https://doi.org/10.1103/PhysRevD.85.014019}{Phys. Rev. D \textbf{85}, 014019 (2012)},
\href{https://arxiv.org/pdf/1111.5921}{[arXiv:1111.5921 [hep-ph]].}
		
\bibitem{NPLQCD:2010ocs}
S.~R.~Beane \textit{et al.} [NPLQCD],
``Evidence for a Bound H-dibaryon from Lattice QCD,''
\href{doi:10.1103/PhysRevLett.106.162001}{Phys. Rev. Lett. \textbf{106}, 162001 (2011)},
\href{http://www1.jlab.org/Ul/publications/view_pub.cfm?pub_id=10154}{[arXiv:1012.3812 [hep-lat]].}

\bibitem{NPLQCD:2011naw}
S.~R.~Beane \textit{et al.} [NPLQCD],
``The Deuteron and Exotic Two-Body Bound States from Lattice QCD,''
\href{https://doi.org/10.1103/PhysRevD.85.054511}{Phys. Rev. D \textbf{85}, 054511 (2012)},
\href{https://arxiv.org/pdf/1109.2889}{[arXiv:1109.2889 [hep-lat]].}

\bibitem{Ishii:2006ec}
N.~Ishii, S.~Aoki and T.~Hatsuda,
``The Nuclear Force from Lattice QCD,''
\href{https://doi.org/10.1103/PhysRevLett.99.022001}{Phys. Rev. Lett. \textbf{99}, 022001 (2007)},
\href{https://arxiv.org/pdf/nucl-th/0611096}{[arXiv:nucl-th/0611096 [nucl-th]].}
	
	
\bibitem{Inoue:2011nq}
T.~Inoue [HAL QCD],
``Bound H-dibaryon from Full QCD Simulations on the Lattice,''
\href{https://doi.org/10.22323/1.139.0124}{PoS \textbf{LATTICE2011}, 124 (2011)},
\href{https://arxiv.org/pdf/1111.5098}{[arXiv:1111.5098 [hep-lat]].}
	
		
\bibitem{Haidenbauer:2011za}
J.~Haidenbauer and U.~G.~Meissner,
``Exotic bound states of two baryons in light of chiral effective field theory,''
\href{https://doi:10.1016/j.nuclphysa.2012.01.021}{Nucl. Phys. A \textbf{881}, 44-61 (2012)},
\href{https://arxiv.org/pdf/1111.4069}{[arXiv:1111.4069 [nucl-th]].}

		
		
\bibitem{HALQCD:2014okw}
F.~Etminan \textit{et al.} [HAL QCD],
``Spin-2 $N\Omega$ dibaryon from Lattice QCD,''
\href{https://doi.org/10.1016/j.nuclphysa.2014.05.014}{Nucl. Phys. A \textbf{928}, 89-98 (2014)},
\href{https://arxiv.org/pdf/1403.7284}{[arXiv:1403.7284 [hep-lat]].}
		
		
\bibitem{Gongyo:2017fjb}
S.~Gongyo, K.~Sasaki, S.~Aoki, T.~Doi, T.~Hatsuda, Y.~Ikeda, T.~Inoue, T.~Iritani, N.~Ishii and T.~Miyamoto, \textit{et al.}
``Most Strange Dibaryon from Lattice QCD,''
\href{https://doi:10.1103/PhysRevLett.120.212001}{Phys. Rev. Lett. \textbf{120}, no.21, 212001 (2018)},
\href{https://arxiv.org/pdf/1709.00654}{[arXiv:1709.00654 [hep-lat]].}
		
		
\bibitem{Morita:2019rph}
K.~Morita, S.~Gongyo, T.~Hatsuda, T.~Hyodo, Y.~Kamiya and A.~Ohnishi,
``Probing $\Omega\Omega$ and $p\Omega$ dibaryons with femtoscopic correlations in relativistic heavy-ion collisions,''
\href{https://doi:10.1103/PhysRevC.101.015201}{Phys. Rev. C \textbf{101}, no.1, 015201 (2020)},
\href{https://arxiv.org/pdf/1908.05414}{[arXiv:1908.05414 [nucl-th]].}
	
\bibitem{Chen:2019vdh}
X.~H.~Chen, Q.~N.~Wang, W.~Chen and H.~X.~Chen,
``Exotic $\Omega\Omega$ dibaryon states in a molecular picture,''
\href{https://doi:10.1088/1674-1137/abdfbe}{Chin. Phys. C \textbf{45}, no.4, 041002 (2021)},
\href{https://arxiv.org/pdf/1906.11089}{[arXiv:1906.11089 [hep-ph]].}
		
		
\bibitem{Pu:2024kfh}
J.~Pu, K.~J.~Sun, C.~W.~Ma and L.~W.~Chen,
``Probing the internal structures of p\ensuremath{\Omega} and \ensuremath{\Omega}\ensuremath{\Omega} with their production at the Large Hadron Collider,''
\href{https://doi.org/10.1103/PhysRevC.110.024908}{Phys. Rev. C \textbf{110}, no.2, 024908 (2024)},
\href{https://arxiv.org/pdf/2402.04185}{[arXiv:2402.04185 [hep-ph]].}
		
\bibitem{Junnarkar:2024kwd}
P.~M.~Junnarkar and N.~Mathur,
``Spectrum of two-flavored spin-zero heavy dibaryons in lattice QCD,''
\href{https://doi.org/10.1103/PhysRevD.111.014512}{Phys. Rev. D \textbf{111}, no.1, 1 (2025)},
\href{https://arxiv.org/pdf/2410.08519}{[arXiv:2410.08519 [hep-lat]].}
		
\bibitem{HALQCD:2019wsz}
K.~Sasaki \textit{et al.} [HAL QCD],
``$\Lambda\Lambda$ and N$\Xi$ interactions from lattice QCD near the physical point,''
\href{https://doi.org/10.1016/j.nuclphysa.2020.121737}{Nucl. Phys. A \textbf{998}, 121737 (2020)},
\href{https://arxiv.org/pdf/1912.08630}{[arXiv:1912.08630 [hep-lat]].}
		

\bibitem{Morita:2016auo}
K.~Morita, A.~Ohnishi, F.~Etminan and T.~Hatsuda,
``Probing multistrange dibaryons with proton-omega correlations in high-energy heavy ion collisions,''
\href{https://doi.org/10.1103/PhysRevC.94.031901}{Phys. Rev. C \textbf{94}, no.3, 031901 (2016)},
[erratum: Phys. Rev. C \textbf{100}, no.6, 069902 (2019)]
\href{https://arxiv.org/pdf/1605.06765}{[arXiv:1605.06765 [hep-ph]].}


\bibitem{Haidenbauer:2017sws}
J.~Haidenbauer, S.~Petschauer, N.~Kaiser, U.~G.~Mei\ss{}ner and W.~Weise,
``Scattering of decuplet baryons in chiral effective field theory,''
\href{https://doi.org/10.1140/epjc/s10052-017-5309-4}{Eur. Phys. J. C \textbf{77}, no.11, 760 (2017)},
\href{https://arxiv.org/pdf/1708.08071}{[arXiv:1708.08071 [nucl-th]].}

\bibitem{Garcilazo:2018gkb}
H.~Garcilazo and A.~Valcarce,
``$\Omega d$ bound state,''
\href{https://doi.org/10.1103/PhysRevC.98.024002}{Phys. Rev. C \textbf{98}, no.2, 024002 (2018)},
\href{https://arxiv.org/pdf/1810.04382}{[arXiv:1810.04382 [nucl-th]].}
		
\bibitem{HALQCD:2018qyu}
T.~Iritani \textit{et al.} [HAL QCD],
``$N\Omega$ dibaryon from lattice QCD near the physical point,''
\href{https://doi.org/10.1016/j.physletb.2019.03.050}{Phys. Lett. B \textbf{792}, 284-289 (2019)},
\href{https://arxiv.org/pdf/1810.03416}{[arXiv:1810.03416 [hep-lat]].}
		
		
\bibitem{Chen:2021hxs}
X.~H.~Chen, Q.~N.~Wang, W.~Chen and H.~X.~Chen,
``Mass spectra of $N\Omega$  dibaryons in the $^3S_1$ and $^5S_2$ channels,''
\href{https://doi.org/10.1103/PhysRevD.103.094011}{Phys. Rev. D \textbf{103}, no.9, 094011 (2021)},
\href{https://arxiv.org/pdf/1908.05414}{[arXiv:2103.09739 [hep-ph]].}
		
		
\bibitem{Wan:2019ake}
B.~D.~Wan, L.~Tang and C.~F.~Qiao,
``Hidden-bottom and -charm hexaquark states in QCD sum rules,''
\href{https://doi.org/10.1140/epjc/s10052-020-7701-8}{Eur. Phys. J. C \textbf{80}, no.2, 121 (2020)},
\href{https://arxiv.org/pdf/1912.12046}{[arXiv:1912.12046 [hep-ph]].}
		
		
\bibitem{Wang:2024riu}
B.~Wang, K.~Chen, L.~Meng and S.~L.~Zhu,
``Spectrum of molecular hexaquarks,''
\href{https://doi.org/10.1103/PhysRevD.110.014038}{Phys. Rev. D \textbf{110}, no.1, 014038 (2024)},
\href{https://arxiv.org/pdf/2406.06993}{[arXiv:2406.06993 [hep-ph]].}
		
\bibitem{Geng:2024dpk}
Y.~Geng, L.~Liu, P.~Sun, J.~J.~Wu, H.~Xing, Z.~Yan and R.~Zhu,
``Doubly Charmed $H$-like dibaryon $\Lambda_c \Lambda_c$ scattering from Lattice QCD,''
\href{https://doi.org/10.22323/1.466.0307}{PoS \textbf{LATTICE2024}, 307 (2025)},
		
\bibitem{An:2025rjv}
H.~T.~An, S.~Q.~Luo and X.~Liu,
``Doubly-charmed hexaquarks in the diquark picture,''
\href{https://arxiv.org/pdf/2504.06107}{[arXiv:2504.06107 [hep-ph]].}
			
\bibitem{Lee:2011rka}
N.~Lee, Z.~G.~Luo, X.~L.~Chen and S.~L.~Zhu,
``Possible Deuteron-like Molecular States Composed of Heavy Baryons,''
\href{https://doi.org/10.1103/PhysRevD.84.014031}{Phys. Rev. D \textbf{84}, 014031 (2011)},
\href{https://arxiv.org/pdf/1104.4257}{[arXiv:1104.4257 [hep-ph]].}

		
\bibitem{Shiri:2023fjh}
N.~Shiri, N.~Tazimi and M.~Monemzadeh,
``Bound state solutions of the Schr{\"o}dinger equation for dibaryons via asymptotic iteration method,''
\href{https://arxiv.org/pdf/2305.02716}{[arXiv:2305.02716 [hep-ph]].}
		
		
\bibitem{Shah:2024thr}
Z.~Shah, D.~Rathaud and A.~K.~Rai,
``Masses of dibaryonic~$\Xi^*_c$ $\Xi^*_c$ states,''
\href{https://doi.org/10.22323/1.462.0080}{PoS \textbf{HQL2023}, 080 (2024)},
		
		

\bibitem{Richard:2020zxb}
J.~M.~Richard, A.~Valcarce and J.~Vijande,
``Very heavy flavored dibaryons,''
\href{https://doi.org/10.1103/PhysRevLett.124.212001}{Phys. Rev. Lett. \textbf{124}, no.21, 212001 (2020)},
\href{https://arxiv.org/pdf/2005.06894}{[arXiv:2005.06894 [hep-ph]].}
		
		
		
\bibitem{Lyu:2022tsd}
Y.~Lyu, H.~Tong, T.~Sugiura, S.~Aoki, T.~Doi, T.~Hatsuda, J.~Meng and T.~Miyamoto,
``Optimized two-baryon operators in lattice QCD,''
\href{https://doi.org/10.1103/PhysRevD.105.074512}{Phys. Rev. D \textbf{105}, no.7, 074512 (2022)},
\href{https://arxiv.org/pdf/2201.02782}{[arXiv:2201.02782 [hep-lat]].}
		
		
\bibitem{Mathur:2022ovu}
N.~Mathur, M.~Padmanath and D.~Chakraborty,
``Strongly Bound Dibaryon with Maximal Beauty Flavor from Lattice QCD,''
\href{https://doi.org/10.1103/PhysRevLett.130.111901}{Phys. Rev. Lett. \textbf{130}, no.11, 111901 (2023)},
\href{https://arxiv.org/pdf/2205.02862}{[arXiv:2205.02862 [hep-lat]].}
		
		
\bibitem{Martin-Higueras:2024qaw}
P.~Mart{\'\i}n-Higueras, D.~R.~Entem, P.~G.~Ortega, J.~Segovia and F.~Fern{\'a}ndez,
``Study of the {\ensuremath{\Omega}}ccc{\ensuremath{\Omega}}ccc and {\ensuremath{\Omega}}bbb{\ensuremath{\Omega}}bbb dibaryons in a constituent quark model,''
\href{https://doi.org/10.1103/PhysRevD.111.054002}{Phys. Rev. D \textbf{111}, no.5, 054002 (2025)},
\href{https://arxiv.org/pdf/2406.19208}{[arXiv:2406.19208 [hep-ph]].}
		
			
		
\bibitem{Azizi:2019xla}
K.~Azizi, S.~S.~Agaev and H.~Sundu,
``The Scalar Hexaquark $uuddss$: a Candidate to Dark Matter?,''
\href{https://doi.org/10.1088/1361-6471/ab9a0e}{J. Phys. G \textbf{47}, no.9, 095001 (2020)},
\href{https://arxiv.org/pdf/1904.09913}{[arXiv:1904.09913 [hep-ph]].}
		
		
\bibitem{Huang:2013nba}
H.~Huang, J.~Ping and F.~Wang,
``Dynamical calculation of the $\Delta\Delta$ dibaryon candidates,''
\href{https://doi.org/10.1103/PhysRevC.89.034001}{Phys. Rev. C \textbf{89}, no.3, 034001 (2014)},
\href{https://arxiv.org/pdf/1312.7756}{[arXiv:1312.7756 [hep-ph]].}
	

\bibitem{Mutuk:2022zgn}
H.~Mutuk and K.~Azizi,
``Investigation of {\ensuremath{\Delta}}0{\ensuremath{\Delta}}0 dibaryon in QCD,''
\href{https://doi.org/10.1103/PhysRevD.105.094021}{Phys. Rev. D \textbf{105}, no.9, 094021 (2022)},
\href{https://arxiv.org/pdf/2204.03050}{[arXiv:2204.03050 [hep-ph]].}
		
		
\bibitem{Barsbay:2022gtu}
B.~Barsbay, K.~Azizi and H.~Sundu,
``Heavy-light hybrid mesons with different spin-parities,''
\href{https://doi.org/10.1140/epjc/s10052-022-11053-x}{Eur. Phys. J. C \textbf{82}, no.12, 1086 (2022)},
[erratum: Eur. Phys. J. C \textbf{82}, no.12, 1138 (2022)]
\href{https://arxiv.org/pdf/2205.14597}{[arXiv:2205.14597 [hep-ph]].}
		
		

		
\bibitem{Belyaev:1982sa}
V.~M.~Belyaev and B.~L.~Ioffe,
``Determination of Baryon and Baryonic Resonance Masses from QCD Sum Rules. 1. Nonstrange Baryons,''
\href{http://www.jetp.ras.ru/cgi-bin/dn/e_056_03_0493.pdf}{Sov. Phys. JETP \textbf{56}, 493-501 (1982)},
ITEP-59-1982.

		
\bibitem{Belyaev:1982cd}
V.~M.~Belyaev and B.~L.~Ioffe,
``Determination of the baryon mass and baryon resonances from the quantum-chromodynamics sum rule. Strange baryons,''
\href{http://www.jetp.ras.ru/cgi-bin/e/index/e/57/4/p716?a=list}{Sov. Phys. JETP \textbf{57}, 716-721 (1983)},
ITEP-132-1982.
		
\bibitem{Furstenau:1994ra}
H.~Furstenau,
``Scaling violation and fragmentation functions in e+ e- annihilation,''
\href{https://lib-extopc.kek.jp/preprints/PDF/2000/0030/0030725.pdf}{CERN-PPE-94-155.}

\bibitem{Krasnikov:1982ea}
N.~V.~Krasnikov, A.~A.~Pivovarov and N.~N.~Tavkhelidze,
``The Use of Finite Energy Sum Rules for the Description of the Hadronic Properties of QCD,''
\href{https://doi.org/10.1007/BF01577186}{Z. Phys. C \textbf{19}, 301 (1983)}.

\bibitem{Groote:1999zp}
S.~Groote, J.~G.~Korner and A.~A.~Pivovarov,
``O(alpha(s)) corrections to the correlator of finite mass baryon currents,''
\href{https://doi.org/10.1103/PhysRevD.61.071501}{Phys. Rev. D \textbf{61}, 071501 (2000)}
\href{https://arxiv.org/pdf/hep-ph/9911393}{[arXiv:hep-ph/9911393 [hep-ph]].}



\bibitem{Groote:2014pva}
S.~Groote, J.~G.~K{\"o}rner and D.~Niinepuu,
``Perturbative $O(\alpha_s)$ corrections to the correlation functions of light tetraquark currents,''
\href{https://doi.org/10.1103/PhysRevD.90.054028}{Phys. Rev. D \textbf{90}, no.5, 054028 (2014)},
\href{https://arxiv.org/pdf/1401.4801}{[arXiv:1401.4801 [hep-ph]].}



\bibitem{Guo:2017jvc}
F.~K.~Guo, C.~Hanhart, U.~G.~Mei{\ss}ner, Q.~Wang, Q.~Zhao and B.~S.~Zou,
``Hadronic molecules,''
\href{https://doi.org/10.1103/RevModPhys.90.015004}{Rev. Mod. Phys. \textbf{90}, no.1, 015004 (2018)},
[erratum: Rev. Mod. Phys. \textbf{94}, no.2, 029901 (2022)]
\href{https://arxiv.org/pdf/1705.00141}{[arXiv:1705.00141 [hep-ph]].}
		

		
		
\end{thebibliography}
\end{document}